# Multiplexed Supercell Metasurface Design and Optimization with Tandem Residual Networks

*Christopher Yeung[1,2], Ju-Ming Tsai[1], Brian King[1], Benjamin Pham[1], David Ho[1], Julia Liang[1], Mark W. Knight[2], and Aaswath P. Raman[1,*]*

[1]Department of Materials Science and Engineering, University of California, Los Angeles, CA 90095, USA

[2]Northrop Grumman Corporation, Redondo Beach, CA 90278, USA

*Corresponding Author: aaswath@ucla.edu

**Abstract:** Complex nanophotonic structures hold the potential to deliver exquisitely tailored optical responses for a range of applications. Metal-insulator-metal (MIM) metasurfaces arranged in supercells, for instance, can be tailored by geometry and material choice to exhibit a variety of absorption properties and resonant wavelengths. With this flexibility, however, comes a vast space of design possibilities that classical design paradigms struggle to effectively navigate. To overcome this challenge, here we demonstrate a tandem residual network approach to efficiently generate multiplexed supercells through inverse design. By using a training dataset with several thousand full-wave electromagnetic simulations in a design space of over three trillion possible designs, the deep learning model can accurately generate a wide range of complex supercell designs given a spectral target. Beyond inverse design, the presented approach can also be used to explore the structure-property relationships of broadband absorption and emission in such supercell configurations. Thus, this study demonstrates the feasibility of high-dimensional supercell inverse design with deep neural networks that is applicable to complex nanophotonic structures composed of multiple subunit elements that may exhibit coupling.

# 1. Introduction

Nanophotonic materials, including metasurfaces and metamaterials, have greatly expanded our ability to tailor light-matter interaction and deliver new functionalities for information processing and sensing applications [1], [2], [3], [4]. As demand for advanced capabilities and high-performance nanophotonic devices grow, multimodal implementations with interconnected ensembles of optical sub-components, including supercells, have shown great promise in delivering tailored responses with respect to many optical characteristics [5], [6], [7], [8]. For example, complex spatial arrangements within photonic crystal circuits have yielded high-efficiency spatial mode conversion [9]. Similarly, by employing metasurfaces that contain periodic arrays of meta-atoms with different geometric parameters, a range of useful behaviors including out-of-plane beam deflection and mirroring can be demonstrated [10]. Although the incorporation of numerous distinct subunit elements within a photonic structure is desirable, it is accompanied by an exponential increase in design costs as a result of the increased dimensionality of the associated design space [11].

A particular category of periodic metasurface structures that has shown promise in supercell configurations is the metal-insulator-metal (MIM) metasurface absorber. Periodic MIM absorbers yield strong resonances that are narrowband in nature, where the wavelength of the resonance peak can be shifted by changing the shape of the resonator [34], [35], [36], [37]. By adopting simple supercell configurations, which contain more than one resonator geometry, multi-resonant and broadband absorption behavior has previously been realized [38], [39], [40]. The design and optimization of more complex supercells with hybridized behavior, however, remains an open challenge, but holds the potential of yielding a broader range of spectral responses than previously achieved.

Conventional design processes for periodic and complex supercell metasurfaces rely on electromagnetic (EM) simulations that are iteratively optimized by tuning key design parameters until the desired optical properties are obtained. Techniques that have been employed include evolutionary algorithms [12], topology optimization [13], [14], [15], and adjoint-based methods [16], [17]. In the context of supercells and complex/non-periodic arrangements, methods such as Schur complement domain decomposition and overlapping-domain approximation have yielded compelling results [18], [19]. As the unit cell of a metasurface increases in size and complexity, however, computation times from iterative optimization can rapidly escalate from hours to potentially days or weeks. Additionally, optimizations must be repeated and reconfigured for every new target, thus requiring a substantial amount of computational resources and, oftentimes, prior intuition on the capability of a particular class of nanophotonic structures. These computational costs are further compounded by the fact that only the final optimized results are preserved; any prior data generated in an optimization cycle is not typically reused in the future [41]. As a result, iterative design methods also become increasingly inefficient over time [20].

In response to the need for more efficient design strategies, data-driven approaches based on machine learning, such as deep neural networks (DNNs), have found applications in nanophotonic design [21]. DNNs are now well established in many fields, including: natural language processing, drug discovery, materials design, and medical diagnosis [22], [23], [24]. In the photonics context, DNNs have shown promise in designing a diverse range of high-performance structures by directly predicting key geometric parameters (*e.g.*, resonator widths, lengths, radii, etc.). By leveraging a one-time investment of EM simulation training data, DNNs can generate designs at orders-of-magnitude faster speeds than traditional optimization algorithms. An accurate DNN can also be paired with numerical optimization methods to save simulation time, where the DNN identifies solutions near the global minimum and the optimization refines the performance further [30]. With training datasets ranging from several hundred [45] to several thousand [41] instances, previously explored machine learning and DNN-based photonics design include the forward and inverse modeling of multi-shell nanoparticles, multi-layer thin-films, and various classes of metasurfaces [25], [26], [27], [28], [29]. A forward-modeling DNN takes structural parameters as inputs and predicts optical properties such as the absorption spectra. In contrast, an inverse-modeling DNN accepts target optical properties as inputs and generates matching structural parameters. Further advancements in DNNs have led to the development of the tandem network, which is designed to overcome the nonuniqueness scattering problem [27], [41]. However, prior tandem networks have relied on traditional fully connected or dense networks, while tandem implementations with recent architectural advancements such as residual networks or ResNets (which address the well-known vanishing gradient problem [46]) remain unexplored.

While promising, prior studies of DNNs for nanophotonic design have primarily focused on individual scatterers or periodic structures with single-unit cell elements and relatively narrowband operation [31], [32], [33]. Recent studies involving the design of structures with multiple optical elements have assumed that they are separately constructed and then assembled into a multi-element structure [47] [48]. This approach makes the limiting assumption that the coupling between adjacent elements are sufficiently weak, and further does not affect their cross-section. Moreover, in studies where coupling cannot be neglected, separately-trained models were required in order to design metasurfaces with specific numbers of elements [48], which limits scalability. Other work with unit cells consisting of multiple neighboring elements do not solve the inverse problem, but instead develop a fast and accurate proxy or surrogate model for forward design [49]. Therefore, an ML-based strategy for complex supercells that: 1) directly solves the inverse design problem, 2) generates structures with a wide range of unique elements, and 3) considers strong coupling interactions or mode hybridization between individual elements is lacking today, but could allow for the demonstration of complex nanophotonic architectures with a broader range of spectral responses.

In this article, we investigate the inverse design of large multiplexed supercell metasurfaces with over 100 subunit elements that can achieve a diverse set of broadband spectral responses. Specifically, we focus on engineering arbitrary bandwidth absorbers operating in the mid- and long-wave infrared regime (4-12 µm) by designing supercell MIM metamaterial absorbers through a deep learning approach. To navigate the large design space that comes with the increased dimensionality of supercells and to address the vanishing gradient problem associated with deep network architectures, we employed a tandem residual network (shown conceptually in Figure 1A). We demonstrate that with a training dataset of several thousand simulations, in a high-dimensional design space with over three trillion possible design combinations, the network can successfully design narrowband, multi-resonance, and broadband-absorption supercell metasurfaces with high degrees of accuracy. Furthermore, we show that the network itself can be harnessed to approximate the structure-property relationships of the explored class of metasurfaces.

## 2. Results and discussion
### 2.1 Data preparation for deep learning

Nanophotonic supercell structures such as MIM metasurfaces are capable of producing unique optical responses that extend beyond the sum of their parts. Specifically, in addition to the superposition of individual responses, distinct responses may also arise from the interaction or hybridization between neighboring elements [50]. Several examples of such interactions are presented in Figure S1, where the absorption spectra and EM field profiles of various supercell designs are shown. In this figure, we show that specific arrangements of identical subunit resonators can yield absorption peaks with different amplitudes and wavelengths, or new peaks entirely, in comparison to the response of the individual elements. These examples reveal that the relative positions of individual elements are critical as the characteristics of the peaks can depend strongly on which resonators are adjacent to each other. Therefore, supercell-class metasurfaces can potentially access new domains of functionalities by leveraging multiscale optical phenomena. To this end, a comprehensive inverse design scheme for supercell structures must consider the geometries and physical arrangements of individually-integrated structures as well as their collective EM interactions.

Figure 1B presents the detailed implementation of our supercell inverse design strategy. First, we defined a supercell layout of MIM resonators (labeled "1" in Figure 1B). The layout contains an assortment of 100 nm-thick gold cross-shaped resonators with a 100 nm gold backing and 200 nm $Al_2O_3$ spacer. This class of metasurfaces was derived from existing literature on selective thermal emitters and exhibits narrowband resonances in the mid-infrared (MIR) range [42]. A single supercell design is represented by an array of cross-shaped resonator lengths (ranging from 1.4-3 µm in 0.2 µm steps), each with fixed widths (500 nm). The resonator arrays (labeled "2" in Figure 1B) embody a quadrant of the supercell and resembles a hexagonal close-packed (HCP) lattice with a twin boundary, where the individual resonators are mirrored along the diagonal plane. The quadrant is then mirrored along the x- and y-axes to create a four-fold symmetric supercell. The HCP configuration is designed to maximize the area density (and therefore the resonance efficiency) of the supercell, while the four-fold symmetry ensures the structure is s- and p-polarization independent under normal incidence. We limited our supercell size to 25 unique resonators per quadrant (12.8 × 12.8 µm$^2$ before four-fold symmetry) to maximize the resonance modes within the 4-12 µm window while simultaneously attempting to minimize simulation time. Thus, a unique supercell design is represented by $D_A=[l_1, l_2, …, l_{25}]$, with $l_{25}$ being the length of the *25*-th resonator (where $D_A, …, D_n$ are vectors with distinct *l*-values). These vectors were then used as the supercell design parameters for deep learning.

We converted the supercell design parameters into three-dimensional MIM structure models (labeled "3" in Figure 1B) and performed full-wave EM simulations (Lumerical FDTD) on these models over the spectral range of 4-12 µm at normal incidence (labeled "4" in Figure 1B), obtaining an 800-point "ground truth" absorption spectrum for each structure (labeled "5" in Figure 1B). Using this approach, we simulated the absorption spectrum (*A*) for pseudo-randomly generated design parameters (*D*) to create training data pairs (*D*, *A*) for the neural network. As discussed in the next section, the deep learning model "learns" by comparing the ground truth spectra (*A*) to the network-predicted spectra (*A′*, labeled "6" in Figure 1B).

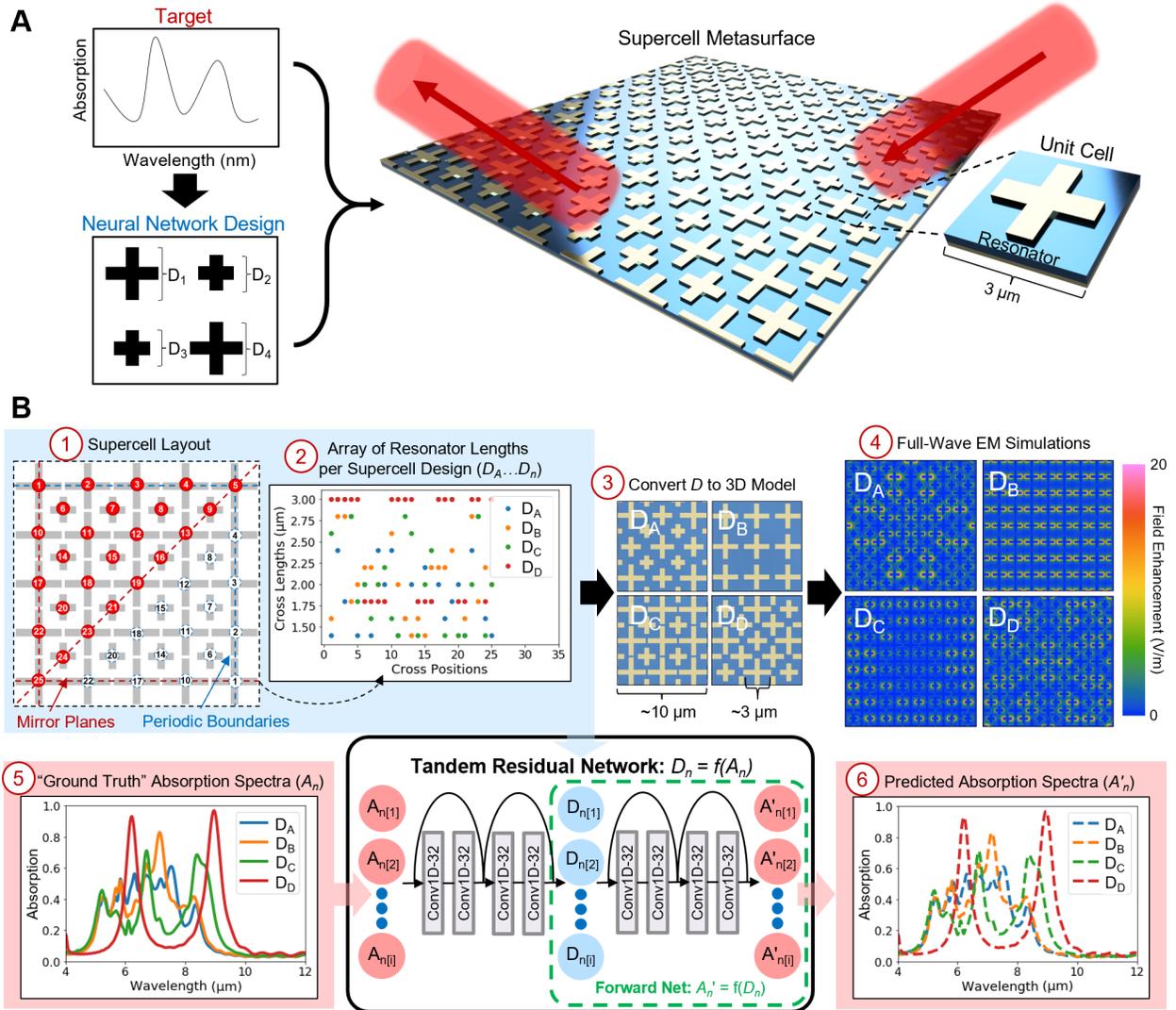

**Figure 1.** Inverse design of supercell metasurfaces with a range of underlying symmetries using a tandem residual network approach. (A) A target absorption spectrum is defined, and the matching design parameters for a multiplexed array of MIM resonators are generated. (B) Data preparation schematic for deep learning. Supercell design parameters (1) representing resonator lengths and positions (2) are converted into 3D models (3). Full-wave electromagnetic simulations are performed on the models (4). Design parameters along with corresponding "ground truth" (5) and predicted (6) absorption spectra are used to train the tandem residual network.

## 2.2 Network characterization and evaluation

The performance of a tandem network hinges on the accuracy of the forward-modeling network as well as the breadth and size of the training dataset. Thus, we sought to optimize the architecture of the forward-modeling network and to ensure that the size of our training dataset maximizes the network's implementation efficiency and predictive capabilities. Unlike previous implementations of the tandem architecture, our approach utilizes one-dimensional convolutional neural networks (1-D CNNs) instead of dense networks. 1-D CNNs have been used in various scientific domains [51] [52], with recent works showing that they are capable of outperforming dense networks in terms of regression fidelity and generalization capabilities [53]. This is enabled by the convolutional layers of the CNN, which are optimized to extract highly discriminative features using a large set of 1-D filter kernels [51]. Furthermore, our particular CNN consists of residual building blocks, which leverage identity shortcuts or skipped connections to address the vanishing gradient problem and achieve better performance than "plain" networks of the same depth [54] [55] [56]. The corresponding ResNet was trained in the forward-modeling configuration to predict an absorption spectrum ($A'$), given a set of design parameters ($D$) as inputs. We evaluated and compared the performances of the dense network, CNN, and ResNet in Figure S2 and S3, where it can be observed that the ResNet achieved the lowest validation loss out of the three model types. Our optimized forward-modeling ResNet architecture consists of a 25-neuron input layer (matching the vector size of the supercell design parameters $D$, with values normalized from 0 to 1), two residual blocks, followed by an 800-neuron dense layer. Each residual block contains two 1-D convolutional layers with 32 filters, kernel size of 3, and zero-padding. In addition, the Adam optimizer, batch size of 10, and ReLU activation functions yielded the lowest validation loss.

Using the same hyperparameters as the forward network, except with an inverted sequence of input and output layers (with 800 and 25 filters, respectively), we designed an inverse-modeling network for the prediction of design parameters ($D'$) given an input $A$. However, plain inverse modeling networks are known to encounter the nonuniqueness problem [27], [41], where the multiple mappings between an EM response and its available structural parameters may confound the network's learning process. To illustrate this problem in the context of our training data, Figure S4 shows several examples where two substantially different supercell design layouts map to nearly-identical dual-band and triple-band responses. Due to the considerable degrees of freedom in a supercell design, the nonuniqueness problem in a supercell architecture is exacerbated relative to single-element and periodic structures, and is crucial to address. Thus, to account for this issue, we implemented the tandem architecture by coupling the inverse-modeling network with a pretrained forward-modeling network.

First, we trained a standard tandem dense network by minimizing the loss function between the input absorption spectrum ($A$) and the spectrum predicted by the forward-modeling network ($A'$), where $A'$ is generated by the same $D'$ predicted by the inverse-modeling network from above. As in Ref. [27], we define the tandem network's loss as the mean squared error between $A$ and $A'$: $MSE = \frac{1}{n}\Sigma(A'_i - A_i)^2$. This loss calculation is distinct from the plain inverse modeling network, which is programmed to minimize the loss between the designs from the training dataset ($D$) and the predicted designs ($D'$). Since $D'$ may offer a completely different solution than $D$ that correctly maps to the target response (due to the issue of nonuniqueness), a plain inverse-modeling network can struggle to minimize loss or converge, whereas in the tandem network, the loss function converges so long as the target and predicted spectra ($A$ and $A'$) are similar [27], [41]. In other words, since the task at hand requires solving a one-to-many problem, the tandem network finds the optimum response for an input target rather than mixing the corresponding outputs (which can lead to a suboptimal solution).

To validate that the tandem network reaches said optimum response, in Figure S5, we show the tandem neural network architecture's ability to resolve the nonuniqueness issue by testing the accuracy of designs for which an explicit nonunique relationship exists, which were found in Figure S4. As seen in Figure S5A, dual-band and triple-band spectra were passed into the inverse modeling network, and a poor match between the input spectra and the simulated design parameters can be observed. However, when the same spectra were passed into the tandem network (Figure S5B), the accuracy between the input spectra and the simulated parameters is substantially improved. Thus, we find that the tandem dense network effectively addresses the nonuniqueness issue, and yields superior accuracy over a plain inverse modeling network for the multiplexed supercell metasurfaces evaluated here. Furthermore, in Figure S6, we verify that the quantity of our training data was capable of maximizing the network's ability to learn supercell designs.

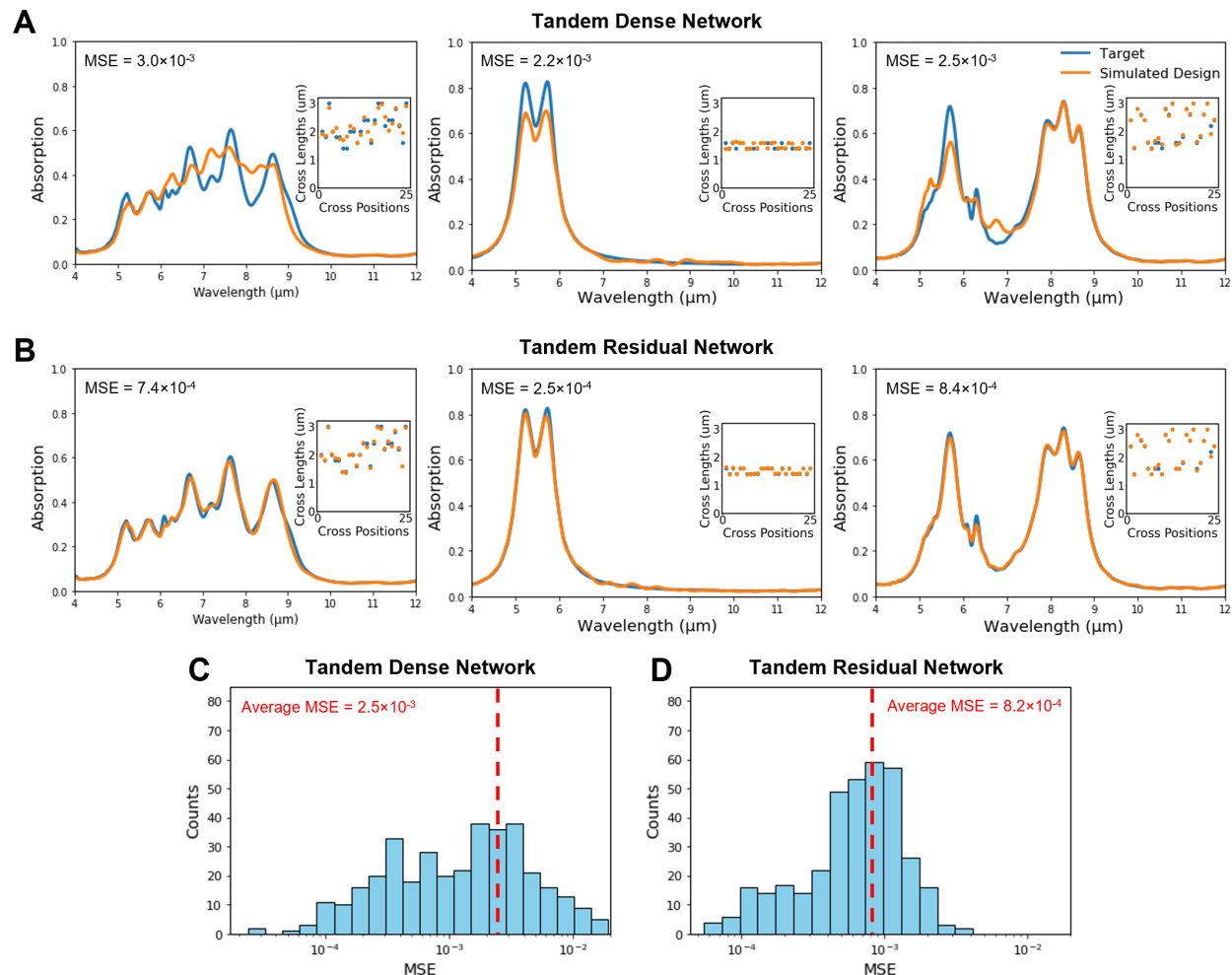

**Figure 2.** Performance evaluation based on mean-squared error (MSE) of target spectra and simulated designs for (A) the tandem dense and (B) tandem residual networks. The tandem residual network achieves a lower MSE for three distinct targets. Inset images show the design parameters for the corresponding spectra. Statistical analyses across the entire validation dataset (over 300 spectra) for the tandem dense (C) and residual (D) networks indicate that the latter achieves a lower average MSE.

Using the customized loss function approach described above, we implemented a tandem residual network and compared its performance to the tandem dense network. Figure 2 presents several example test results from both networks; with new spectral targets from the validation dataset. These test results are obtained by simulating the predicted $D'$, then comparing the target and simulated spectra (shown as blue and orange lines, respectively). It can be observed that across various spectral response patterns, the tandem residual network (Figure 2B) produces supercell designs with greater accuracy than the tandem dense network (Figure 2A). A larger statistical evaluation using the entire validation dataset (360 input spectra) reveals that the average validation MSE is approximately $2.5 \times 10^{-3}$ for the tandem dense network (Figure 2C) and $8.2 \times 10^{-4}$ for the tandem residual network (Figure 2D). Moreover, we observe that the tandem residual network exhibits a lower overall distribution of errors. As a result, we demonstrate that a tandem architecture composed of 1-D convolutional layers and residual building blocks is well-suited for supercell design and can outperform a tandem network (of the same network depth) that is based on fully connected layers.

### 2.3 Inverse design of multi-resonance and broadband metasurfaces

We utilized the tandem residual network to generate new supercell metasurface designs with a broad range of spectral properties. Figure 3 presents a series of test cases comparing the target network inputs to the simulated results of the corresponding output designs. The inset images show the spatial geometries of each supercell designed by the network. For example, as shown in Figure 3A, after specifying a narrowband target with a full width half maximum (FWHM) of 0.5 µm, the network generated a periodic layout that matches the target spectra with over 90% accuracy as well as results from prior literature [42]. Similarly, in Figure 3B and 3C, dual-narrowband and triple-narrowband designs were created (with sharp resonances at two and three discrete wavelengths) that closely match their respective targets. In these multi-resonance structures, the supercells include additional cross dimensions that are associated with distinct resonances.

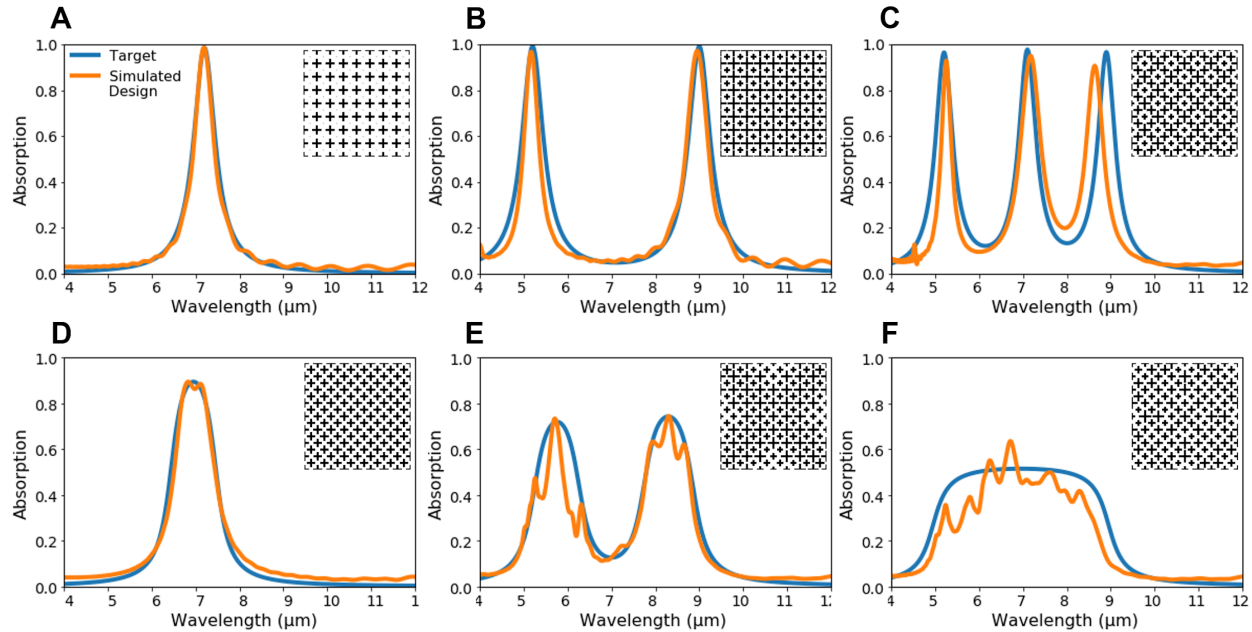

**Figure 3.** Inverse design of new supercell metasurfaces with the tandem neural network. The structures exhibit (A) narrowband, (B) dual-narrowband, (C) triple-narrowband, (D) broadband, (E) dual-broadband, and (F) graybody behaviors. Blue lines indicate the target spectra used as inputs to the network, and orange lines represent the simulated results of the output design parameters. Inset images show the physical layouts of the network-generated supercells.

The ability to construct an array of resonator geometries suggests that different resonant modes can be superimposed to achieve responses of arbitrary bandwidth [43]. Accordingly, we tasked the neural network with designing metasurfaces with various broadband characteristics (FWHM > 1 µm). In Figure 3D, a broadband structure with a FWHM of 1.5 µm is shown, and in Figure 3E, we increased the complexity of the target to design a structure with dual-broadband absorption peaks. Lastly, in Figure 3F, we demonstrate the design of a broadband graybody structure that encompasses the entire MIR range of resonance wavelengths captured by the training dataset (5-9 µm).

In the design of the aforementioned multi-resonance and broadband structures, the network not only defined the resonator dimensions required to achieve resonances at the target wavelengths, but also determined their appropriate placements within the lattice in order to reach the target absorption amplitudes. For example, as shown in Figure 4A, the network-designed triple-narrowband structure possesses three primary cross lengths that are responsible for resonances at 5.2, 7.2, and 8.6 µm. The high absorption amplitudes are attributed to the periodic and short-range ordered arrangements (repeating patterns spanning 1-2 subunit cell distances) of the resonators, which result in the strong dipole resonances seen in the electric field enhancement plots. When short-range order is converted to long-range order (patterns spanning beyond 2 subunit cell distances), additional response types are enabled. In particular, we observe in Figure 4B that the network can alter the relative positions of the same subunit resonators to produce a new response with lower peak amplitudes and a peak shift. As seen in the EM fields, this response is achieved by the new interactions that emerge from the modified arrangement of resonators, as well as modifications of the cross section of a given resonator by its neighbors. By introducing a larger assortment of cross geometries with more complex interactions, the net absorption spectra can also produce a graybody response (Figure 4C). Thus, by systematically predicting the subunit resonator dimensions as well as their spatial positions, the trained network can modulate absorption peak phase and amplitude by designing multiplexed metasurfaces with a range of underlying symmetries.

**2.4 Estimating the structure-property relationships of the explored metasurface class**

It is a challenging design task to combine multiple distinct resonant modes in a single metastructure while leveraging, or alternatively minimizing, hybridization between modes and maintaining high absorption per unit area [43], [44]. Thus, multiplexed resonator structures impose an inherent tradeoff between broadband response and maximum absorption. Here, we seek to investigate the structure-property relationships of the explored metasurface class by leveraging the near-instantaneous calculation speed of the neural network. Previous studies have used pre-trained ML models for design space exploration and pattern discovery [32] [57] [58] [59]. For instance, it has been shown that for a constrained domain, the fast inference speed of ML models can produce reasonably accurate estimates of an optical system's physical responses so that unnecessary exploration of the solution space can be avoided [60]. Similarly, to enable the exploration of our supercell design space, we use the pre-trained forward-modeling network as a validation mechanism for the design parameters predicted by the tandem network. As one example, we specified design targets using Lorentzian functions of increasing bandwidth (FWHM of 0.2-4 µm centered at 7 µm), illustrated in Figure 5A. The tandem network outputs were then fed into the forward-modeling network, and the resulting design predictions were compared to the initial targets. Full-wave simulation results of the tandem network-designed structures are also presented in Figure 5A, indicating that the network-predicted results match well with the ground truth. In this approach, the forward network effectively serves as a high-speed surrogate EM solver, replacing the FDTD software that was used to generate the training data.

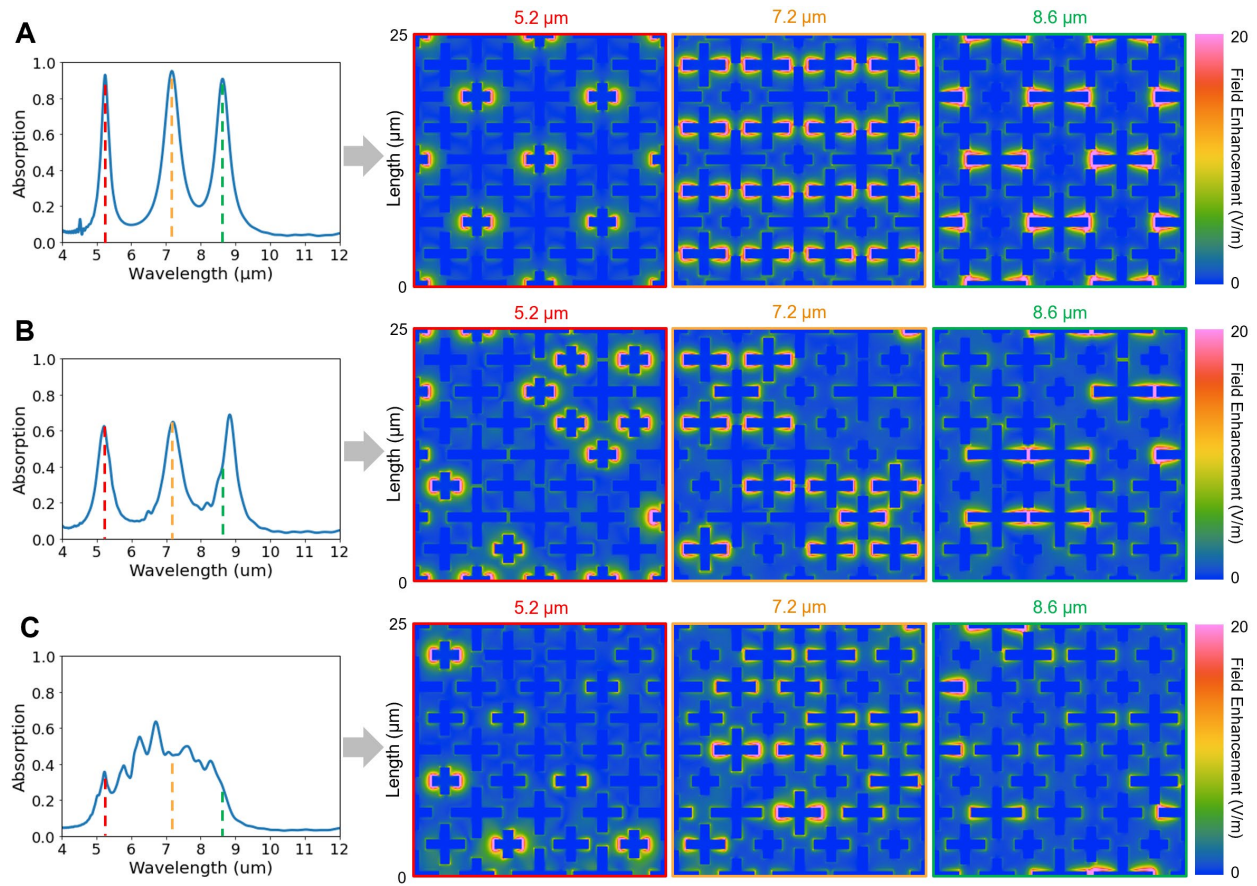

**Figure 4.** Relationships between supercell absorption properties and their subunit resonator spatial distributions. Simulated absorption spectra (of the tandem network-designed structures) and corresponding electric field profiles are shown for triple-narrowband structures with (A) high and (B) low absorption peaks and a (C) graybody structure. These plots reveal the dependence of absorption response on resonator geometry and position relative to other elements.

The design predictions reveal that when an unobtainable target was specified, the network designs a structure with the closest possible solution in the context of the supercell design space which it was trained on. As a result, we observe that as the target bandwidth increases, the discrepancy between the target response and the closest design (measured by MSE) increases as well (Figure 5B). This, in turn, allows us to numerically infer a relationship between the broadband response and the maximum obtainable absorption for this class of resonant metasurfaces, as estimated by the machine learning algorithm. By fitting these observations, we can further derive an estimate for maximum absorption at various bandwidths ($R^2 = 0.98$)

$$A_{max} = 0.0004f^2 - 0.0302f + 1.0154, \qquad (1)$$

where $A_{max}$ is the model's estimate of maximum absorption and $f$ is the FWHM in THz. While we show that the structure-property relationships of a design space can be easily represented using a forward-modeling network, we note that the relation in Eqn. 1 (or any relation captured in a similar manner) is not universally applicable, but subject to the same parametric restrictions that were imposed on the training data (*i.e.*, cross widths fixed at 500 nm and cross lengths within the range of 1.4-3 µm). Metasurfaces with dimensions beyond the range restricted by the training data may exhibit a relationship that is different from Eqn. 1. However, the training dataset may simply be updated to incorporate a wider range of geometries to account for such parametric restrictions.

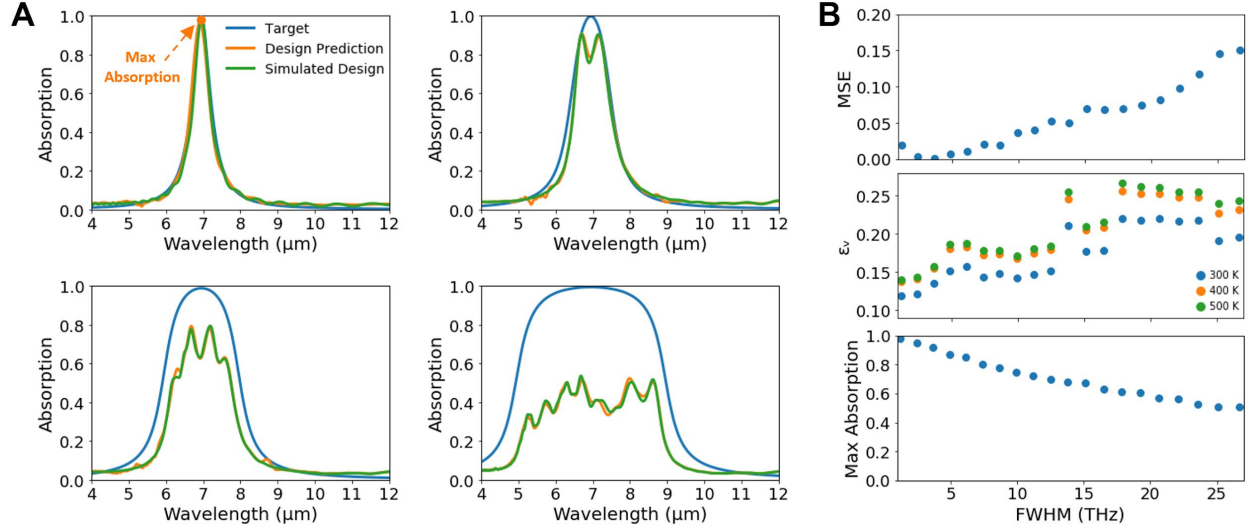

**Figure 5.** Probing metasurface design relations using the tandem network. (A) Tandem network input targets (blue lines) for various Lorentzian functions (FWHM of 0.6, 1, 2, and 3 µm centered at 7 µm) and the corresponding forward network-predicted results (orange lines). The network is unable to identify designs that exceed the model's estimate of bandwidth / maximum absorption of the explored class of supercell metasurfaces. Full-wave simulation results are shown (green lines) for comparison, indicating that the network-predicted results match well with the ground truth. (B) Network-determined design trends and metrics, including the MSE between target and design responses, thermal emittance of the metasurface, and max absorption as functions of FWHM (THz).

As an additional example of discovering application-specific design insights through the neural network, we can calculate the average normal-incidence emissivity of the optimized supercell metasurfaces within defined target bandwidths:

$$\bar{\varepsilon} = \frac{\int_{v_1}^{v_2} I_{BB}(T,v) \cdot \varepsilon(v) dv}{\int_{v_1}^{v_2} I_{BB}(T,v) dv}. \quad (2)$$

Here, $I_{BB}(T,v) = \frac{2hv^3}{c^2} \frac{1}{e^{\frac{hv}{k_B T}}-1}$ is the spectral radiance of a blackbody at temperature $T$, where $h$ is Planck's constant, $k_B$ is the Boltzmann constant, $c$ is the speed of light, and $v$ is frequency. The lower and upper bounds of the integral ($v_1$ and $v_2$) are derived from the evaluated spectral range (4-12 μm). $\varepsilon(v)$ is the metasurface's spectral emittance, which is equal to $A(v)$ by Kirchoff's law. In this case, by querying the neural network in a cyclic manner to solve for emittance (at various temperatures) as a function of the target bandwidth, we can find the relationship between the two parameters (Figure 5B) in a remarkably short time frame (less than one minute). Overall, we observe that as the sought bandwidth (FWHM) increases, the MSE between the target (with maximum absorption across the entire bandwidth) and design response increases and the maximum absorption point of the achievable design decreases. Furthermore, the integrated normal incidence emittance increases as the additional bandwidth compensates for the decreases in the peak absorption/emittance value. However, as can be seen in Figure 5B, the precise relationship is complex and depends both on the bandwidth being specified and the temperature of the metasurface because the blackbody spectral radiance changes with temperature. Thus, by training a neural network that is tasked with the inverse design of complex supercell metasurfaces, we demonstrate that the same framework can be strategically leveraged to rapidly identify design trends and dependencies associated with application-specific properties (within the parameter ranges represented by the trained class of metasurfaces).

## 3. Conclusions

In this article, we demonstrated a machine learning approach to the inverse-design of multiplexed supercell metasurfaces with over 100 subunit elements. The added degrees of freedom offered by a supercell architecture, relative to periodic single-element structures, yields new tailored capabilities including multi-resonant and broadband responses. By forming a cascaded architecture with an inverse-modeling and forward-modeling network, we show that a tandem network effectively overcomes the nonuniqueness problem present in supercell architectures, and can successfully learn a high-dimensional design space of over three trillion possible designs using only 3,600 data instances. Moreover, we present a network architecture based on 1-D convolutional layers and residual building blocks that is capable of generating designs with greater accuracy than a conventional tandem network based on fully connected layers. Through the superposition and coupling of multiple resonant modes in a compact region, the tandem residual network can efficiently design supercell structures with a range of symmetries that yield narrowband, broadband, and multi-resonant responses. The network not only predicts the geometric parameters for an array of resonators (*e.g.*, resonator widths, lengths, radii, etc.), but also their optimum spatial arrangement towards satisfying a specified target. Therefore, the presented approach enables additional degrees of complexity in metasurface design by directly generating structures with a wide range of unique elements while accounting for coupling between these elements. Though we sought to maximize implementation efficiency by minimizing the required training data, we expect that the performance of our tandem network can be improved with more training data and a larger network architecture. Furthermore, we demonstrate that the network itself can be utilized to approximate the structure-property relationships of the investigated class of metasurfaces (within the parameter ranges represented by the training data). By using the forward-modeling network as a full-wave EM simulator, high-speed parameter sweeps can be performed to capture property-specific design trends such as maximum absorption and thermal emittance as a function of bandwidth. Importantly, our results show that DNN-based approaches can efficiently design and characterize large-scale supercell metasurfaces with numerous discrete resonators. We believe our results can expedite the development of supercell-class nanophotonic structures and materials, which may in turn yield new tailored capabilities not achievable through conventional periodic nanostructures.

## 4. Supplementary Material

Included in the supplementary material are details regarding hyperparameter optimization and training data size.

**Acknowledgements:** This work was supported by the Sloan Research Fellowship from the Alfred P. Sloan Foundation.

## References

[1] Olthaus J, Schrinner P, Reiter D. Optimal Photonic Crystal Cavities for Coupling Nanoemitters to Photonic Integrated Circuits. Adv Quantum Technol 2020, 3:1900084.

[2] Yoshimi H, Yamaguchi T, Ota Y, Arakawa Y, Iwamoto S. Slow light waveguides in topological valley photonic crystals. Opt Lett 2020, 45:2648-2651.

[3] Bin Tarik F, Famili A, Lao Y, Ryckman J. D. Robust optical physical unclonable function using disordered photonic integrated circuits. Nanophotonics 2020, 20200049.

[4] Mittapalli V, Khan H. Excitation Schemes of Plasmonic Angular Ring Resonator-Based Band-Pass Filters Using a MIM Waveguide. Photonics 2019, 6(2):41.

[5] Ding F, Wang Z, He S, Shalaev VM, Kildishev AV. Broadband high-efficiency half-wave plate: a supercell-based plasmonic metasurface approach. ACS Nano 2015, 9(4):4111-9.

[6] Aoni RA, Rahmani M, Xu L, et al. High-efficiency visible light manipulation using dielectric Metasurfaces. Sci Rep 2019, 9(1):1-9.

[7] Wu PC, Tsai WY, Chen WT, et al. Versatile polarization generation with an aluminum plasmonic metasurface. Nano Lett 2017, 17(1):445-52.

[8] Ma Q, Chen L, Jing HB, et al. Controllable and programmable nonreciprocity based on detachable digital coding metasurface. Adv Opt Mater 2019, 7(24):1901285.

[9] Liu V, Miller DA, Fan S. Ultra-compact photonic crystal waveguide spatial mode converter and its connection to the optical diode effect. Opt Express 2012, 20(27):28388-97.

[10] Guo X, Ding Y, Chen X, Duan Y, Ni X. Molding Free-Space Light with Guided-Wave-Driven Metasurfaces. 2020, arXiv preprint, arXiv:2001.03001.

[11] Hegde RS. Deep learning: a new tool for photonic nanostructure design. Nanoscale Adv 2020, 2: 1007–1023.

[12] Gondarenko A, Lipson M. Low Modal Volume Dipole-like Dielectric Slab Resonator. Opt Express 2008, 16:17689-17694.

[13] Kao CY, Osher S, Yablonovitch E. Maximizing Band Gaps in Two-Dimensional Photonic Crystals by Using Level Set Methods. Appl Phy. B: Lasers Opt 2 2005, 81:235-244.

[14] Piggott AY, Lu J, Lagoudakis KG, et al. Inverse Design and Demonstration of a Compact and Broadband on-Chip Wavelength Demultiplexer. Nat Photonics 2015, 9(6):374-377.

[15] Shen B, Wang P, Polson R, Menon R. An Integrated Nanophotonics Polarization Beamsplitter with 2.4x2.4 μm2 Footprint. Nat Photonics 2015, 9:378-382.

[16] Oskooi A, Mutapcic A, Noda S, et al. Robust Optimization of Adiabatic Tapers for Coupling to Slow-Light Photonic-Crystal Waveguides. Opt Express 2012, 20:21558-21575.

[17] Seliger P, Mahvash M, Wang C, Levi A. Optimization of Aperiodic Dielectric Structures. J Appl Phys 2006, 100:034310.

[18] Verweij S, Liu V, Fan S. Accelerating simulation of ensembles of locally differing optical structures via a Schur complement domain decomposition. Opt Lett 2014,39(22):6458-61.

[19] Lin Z, Johnson SG. Overlapping domains for topology optimization of large-area metasurfaces. Opt Express 2019, 27(22):32445-53.

[20] Elesin Y, Lazarov BS, Jensen JS, Sigmund O. Time domain topology optimization of 3D nanophotonic devices. Photonic Nanostruct 2014, 12(1):23-33.

[21] Yeung C, Tsai JM, King B, Kawagoe Y, Ho D, Knight M, Raman AP. Elucidating the Behavior of Nanophotonic Structures Through Explainable Machine Learning Algorithms. ACS Photonics 2020.

[22] Abiodun OI, Jantan A, Omolara AE, Dada KV, Mohamed NA, Arshad H. State-of-the-art in artificial neural network applications: A survey. Heliyon 2018, 4(11):e00938.

[23] Muhammad W, Hart GR, Nartowt B, Farrell JJ, Johung K, Liang Y, Deng J. Pancreatic cancer prediction through an artificial neural network. Frontiers in Artificial Intelligence 2019, 2:2.

[24] Conduit B, Jones N, Stone H, Conduit G. Design of a nickel-base superalloy using a neural network. Mat and Des 2017, 131:358-365.

[25] So S, Rho J. Designing nanophotonic structures using conditional deep convolutional generative adversarial networks. Nanophotonics 2019, 8:1255–1261.

[26] Liu Z, Zhu D, Rodrigues SP, Lee KT, Cai W. Generative Model for the Inverse Design of Metasurfaces. Nano Lett 2018, 18:6570–6576.

[27] Liu D, Tan Y, Khoram E, Yu Z. Training deep neural networks for the inverse design of nanophotonic structures. ACS Photonics 2018, 5(4):1365-9.

[28] Peurifoy J, Shen Y, Jing L,et al. Nanophotonic particle simulation and inverse design using artificial neural networks. Sci Adv 2018, 4(6):eaar4206.

[29] An S, Fowler C, Zheng B, et al. A deep learning approach for objective-driven all-dielectric metasurface design. ACS Photonics 2019, 6(12):3196-207.

[30] Hegde R. Photonics Inverse Design: Pairing Deep Neural Networks With Evolutionary Algorithms. IEEE J. Sel. Top. Quantum Electron 2019, 26(1):2933796.

[31] Ma W, Cheng F, Liu Y. Deep-learning-enabled on-demand design of chiral metamaterials. ACS Nano 2018, 12(6):6326-34.

[32] Inampudi S, Mosallaei H. Neural network based design of metagratings. Appl Phys Lett 2018, 112(24):241102.

[33] Harper ES, Coyle EJ, Vernon JP, Mills MS. Inverse design of broadband highly reflective metasurfaces using neural networks. Phys Rev B 2020, 101(19):195104.

[34] Ogawa S, Kimata M. Metal-insulator-metal-based plasmonic metamaterial absorbers at visible and infrared wavelengths: a review. Materials 2018, 11(3):458.

[35] Vorobyev AY, Topkov AN, Gurin OV, Svich VA, Guo C. Enhanced absorption of metals over ultrabroad electromagnetic spectrum. Appl Phys Lett 2009, 95(12):121106.

[36] Ye YQ, Jin Y, He S. Omnidirectional, polarization-insensitive and broadband thin absorber in the terahertz regime. JOSA B 2010, 27(3):498-504.

[37] Chen HH, Su YC, Huang WL, Kuo CY, Tian WC, Chen MJ, Lee SC. A plasmonic infrared photodetector with narrow bandwidth absorption. Appl Phys Lett 2014, 105(2):023109.

[38] Ma Y, Chen Q, Grant J, Saha SC, Khalid A, Cumming DR. A terahertz polarization insensitive dual band metamaterial absorber. Opt Lett 2011, 36(6):945-7.

[39] Shen X, Cui TJ, Zhao J, Ma HF, Jiang WX, Li H. Polarization-independent wide-angle triple-band metamaterial absorber. Opt Express 2011, 19(10):9401-7.

[40] Luo H, Cheng YZ, Gong RZ. Numerical study of metamaterial absorber and extending absorbance bandwidth based on multi-square patches. Eur Phys J B 2011, 81(4):387-92.

[41] Gao L, Li X, Liu D, Wang L, Yu Z. A bidirectional deep neural network for accurate silicon color design. Adv Mater 2019, 31(51):1905467.

[42] Liu X, Tyler T, Starr T, Starr AF, Jokerst NM, Padilla WJ. Taming the blackbody with infrared metamaterials as selective thermal emitters. Phys Rev Lett 2011, 107(4):045901.

[43] Fan RH, Xiong B, Peng RW, Wang M. Constructing metastructures with broadband electromagnetic functionality. Adv Mater 2019, 1904646.

[44] Ma W, Wen Y, Yu X. Broadband metamaterial absorber at mid-infrared using multiplexed cross resonators. Opt Express 2013, 21(25):30724-30.

[45] Jiang J, Sell D, Hoyer S, Hickey J, Yang J, Fan JA. Free-Form Diffractive Metagrating Design Based on Generative Adversarial Networks. ACS Nano 2019, 13(8): 8872-78.

[46] He K, Zhang X, Ren S, Sun J. Deep residual learning for image recognition. In Proc IEEE Conference on Computer Vision and Pattern Recognition 2016, 770-8.

[47] Liu Z, Zhu D, Lee K, Kim A, Raju L, Cai W. Compounding Meta-Atoms into Metamolecules with Hybrid Artificial Intelligence Techniques. Adv Mater 2020, 32:1904790.

[48] Naseri P, Hum S. A Generative Machine Learning-Based Approach for Inverse Design of Multilayer Metasurfaces. 2020, arXiv preprint, arXiv:2008.02074.

[49] Zhelyeznyakov M, Brunton S, Majumdar A. Deep learning to accelerate Maxwell's equations for inverse design of dielectric metasurfaces. 2020, arXiv preprint, arXiv:2008.10632.

[50] Prodan E, Radloff C, Halas N J Nordlander P. A Hybridization Model for the Plasmon Response of Complex Nanostructures. Science 2003, 302: 1089171.

[51] Ince T, Kiranyaz S, Eren L, Askar M, Gabbouj M. Real-Time Motor Fault Detection by 1-D Convolutional Neural Networks. IEEE Trans Ind Electron 2016, 63:7067-75.

[52] Xiao B, Xu Y, Bi X, Zhang J, Ma, X. Heart sounds classification using a novel 1-D convolutional neural network with extremely low parameter consumption. Neurocomputing 2020, 392:153-9.

[53] Chao Q, Tao J, Wei X, Wang Y, Meng L, Liu C. Cavitation intensity recognition for high-speed axial piston pumps using 1-D convolutional neural networks with multi-channel inputs of vibration signals. Alexandria Eng J 2020.

[54] Tahersima M, Kojima K, Koike-Akino, T, et al. Deep Neural Network Inverse Design of Integrated Photonic Power Splitters. Sci Rep 2019, 9:1368.

[55] Sajedian I, Kim J, Rho J. Finding the optical properties of plasmonic structures by image processing using a combination of convolutional neural networks and recurrent neural networks. Microsystems Nanoeng 2019, 5:27.

[56] Jiang J, Fan J A. Multiobjective and categorical global optimization of photonic structures based on ResNet generative neural networks. Nanophotonics 2020.

[57] An S, Zheng B, Shalaginov M, et al. A freeform dielectric metasurface modeling approach based on deep neural networks. 2019, arXiv preprint, arxiV:2001.00121.

[58] Melati D, Grinberg Y, Dezfouli M, et al. Mapping the global design space of nanophotonic components using machine learning pattern recognition. Nat Commun 2019, 10:4775.

[59] Nadell C C, Huang B, Malof, J M, Padilla W J. Deep learning for accelerated all-dielectric metasurface design. Opt Express 2019, 27(20):27523-35.

[60] Ma W, Liu Z, Kudyshev Z, Boltasseva A, Cai W, Liu Y. Deep Learning for the design of photonic structures. Nature Photonics 2020, 1-14.

# Supplementary Material

**Supercell Structure and Design Analysis**

      To illustrate the unique optical responses enabled by supercell structures, Figure S1 presents several examples of the absorption spectra obtained from single-resonator metasurface designs (left column). A diverse set of spectra can be achieved by assembling supercell designs from these individual resonators and varying their spatial configurations (center column). For example, Figure S1A shows a two-resonator system with cross lengths of 1.8 and 2.6 µm which individually possess absorption peaks at 6.2 and 8 µm, respectively. One possible supercell configuration produces a direct superposition of the individual responses (Supercell Design 1). However, another configuration can yield different amplitudes and new responses entirely (Supercell Design 2). Specifically, at the wavelength marked with the dashed red line, it can be observed that Supercell Design 2 has a higher absorption than Supercell Design 1. An analysis of the EM field profiles (right column) at this wavelength reveals that Supercell Design 2 possesses regions of high intensity that are generated by interactions between neighboring resonators. In another example (Figure S1B), three cross-shaped resonators with cross lengths of 1.4, 2.2, and 3 µm are shown with absorption peaks at 5.2, 7.2, and 9 µm, respectively. One possible supercell configuration with these resonators can produce a response that includes a lower peak amplitude at 5.2 µm and a broadened peak at 9 µm (Supercell Design 2), in comparison to the direct superposition (Supercell Design 1), as marked with the dashed red line. In Figure S1C, three cross-shaped resonators with cross lengths of 1.8, 2.6, and 3 µm are utilized to make two different supercell designs: one that is an approximate sum of the individual responses (Supercell Design 1), and another that includes a peak shift at the dashed red line which no longer corresponds to the original peak (Supercell Design 2). Lastly, we note in Figure S1D that supercell designs with identical numbers of resonators which do not couple (or weakly couple) with neighboring elements, which can be viewed as incoherent interactions, can also produce different responses (i.e., changes in amplitude) due to the effect neighboring elements have on the absorption and scattering cross section of each resonator. Thus, even in the absence of strong coupling, the positions of the individual elements in a supercell may in turn yield unique behaviors by changing the effective cross section of the overall structure. These examples show that specific arrangements of subunit resonators can yield absorption peaks with different amplitudes and wavelengths, or new peaks entirely, in comparison to the response of the individual elements.

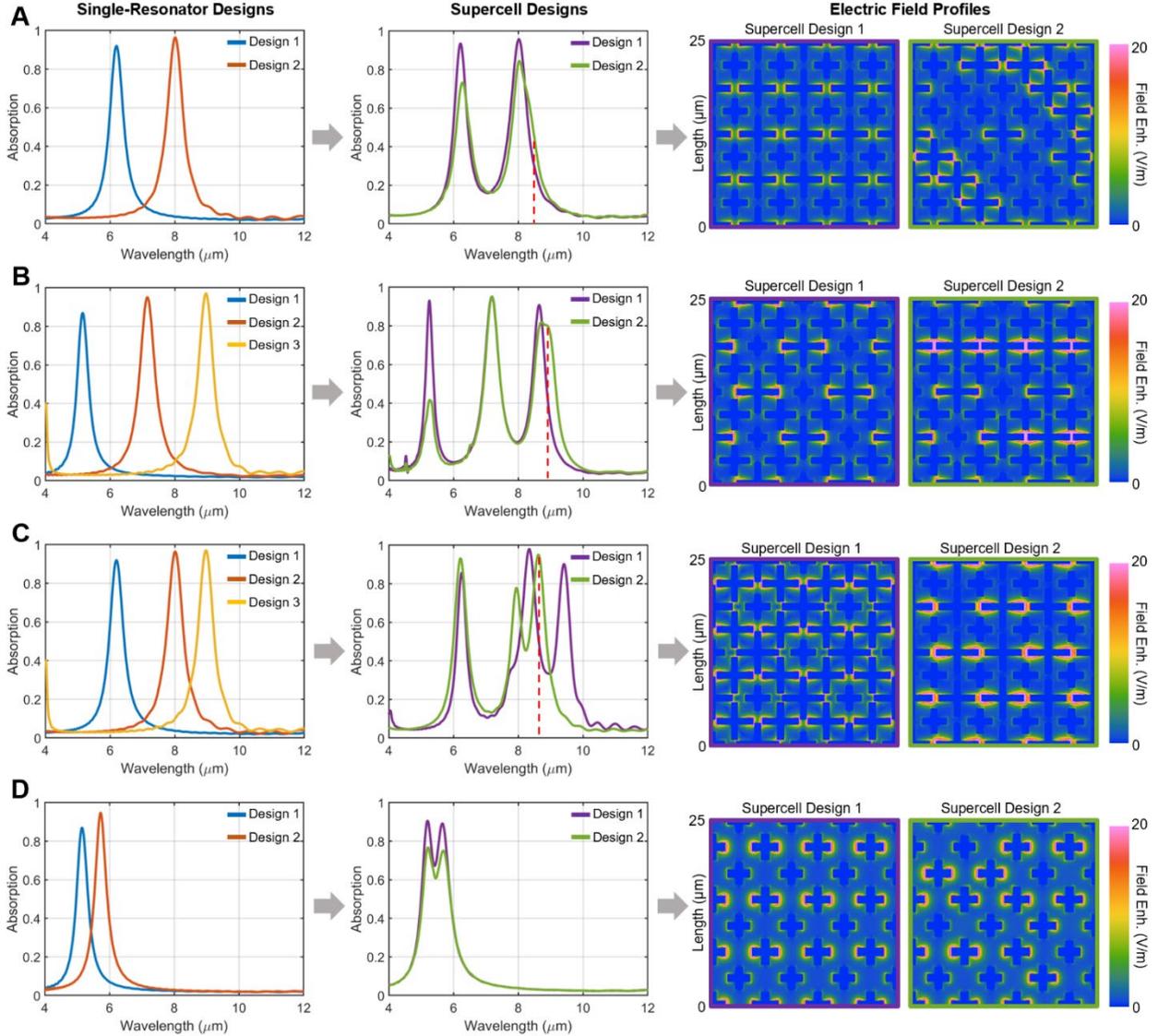

**Figure S1**. Comparison between single-resonator and supercell metasurface designs. The absorption spectra of multiple single-resonator designs (left column) used to create various supercell configurations (center column) that yield distinct optical responses. EM field profiles of the supercell structures (right column), at the wavelengths marked with dashed red lines, reveal interactions between neighboring resonators that result in new absorption responses. (A) Two-resonator and (B) triple-resonator supercells illustrating strong coupling between individual resonators resulting in new peak generation. Additional supercell configurations demonstrating (C) peak shifts and (D) amplitude changes even in the absence of strong coupling due to effects on the absorption cross section.

**Forward-Modeling Network Optimization**

To train the tandem network for inverse design, we first optimized the architecture of the forward-modeling network through extensive hyperparameter tuning. Furthermore, we trained a fully connected or dense network, "plain" 1-D convolutional neural network (CNN), and 1-D residual network (ResNet) to evaluate the highest-performing architecture type. Figure S2 shows the tuning of the dense network, where we compare the validation loss of the starting controlled architecture to the losses of networks trained after changing a single dependent variable. Implemented through the TensorFlow framework, the controlled architecture consists of 2 hidden layers, each with 100 neurons, sigmoid activation functions, a batch size of 10 data instances, and the Adam optimizer. The learning rate is 0.001 with an exponential decay of $10^{-5}$. The tested dependent variables include: number of hidden layers, number of neurons within each layer, activation function, batch size, and optimizer. Figure S2A shows a comparison of different batch sizes (10, 100, and 1000), where we observe noticeable increases in loss as the batch size was increased. In Figure S2B, we tested three commonly used optimization algorithms: Adam, stochastic gradient descent (SGD), and RMSprop. SGD yielded higher losses than the other algorithms while Adam and RMSprop resulted in similar losses. However, RMSprop plateaued much sooner than Adam, indicating that the network was able to improve further with Adam. In Figure S2C, we compared the following activation functions: Sigmoid, TanH (hyperbolic tangent), ReLU (rectified linear unit), Leaky ReLU, and Parametric ReLU. Here, we observe that the Leaky ReLU function resulted in the lowest loss. We then tested various numbers of neurons and layers using the Leaky ReLU activation (Figure S2D and S2E), and found that increasing from 100 to 300 neurons garnered small improvements and faster network convergence, while increasing the number of layers also yielded loss reduction. Figure S2F shows the full integration of the individually-optimized hyperparameters and the use of more elaborate combinations of neurons and layers, which resulted in considerable overall performance improvements. From these tests, we found that the 50-100-200-400 neuron architecture (without batch normalization) has the best performance. Adding more neurons to the optimized architecture did not substantially improve performance and unnecessarily increased training time.

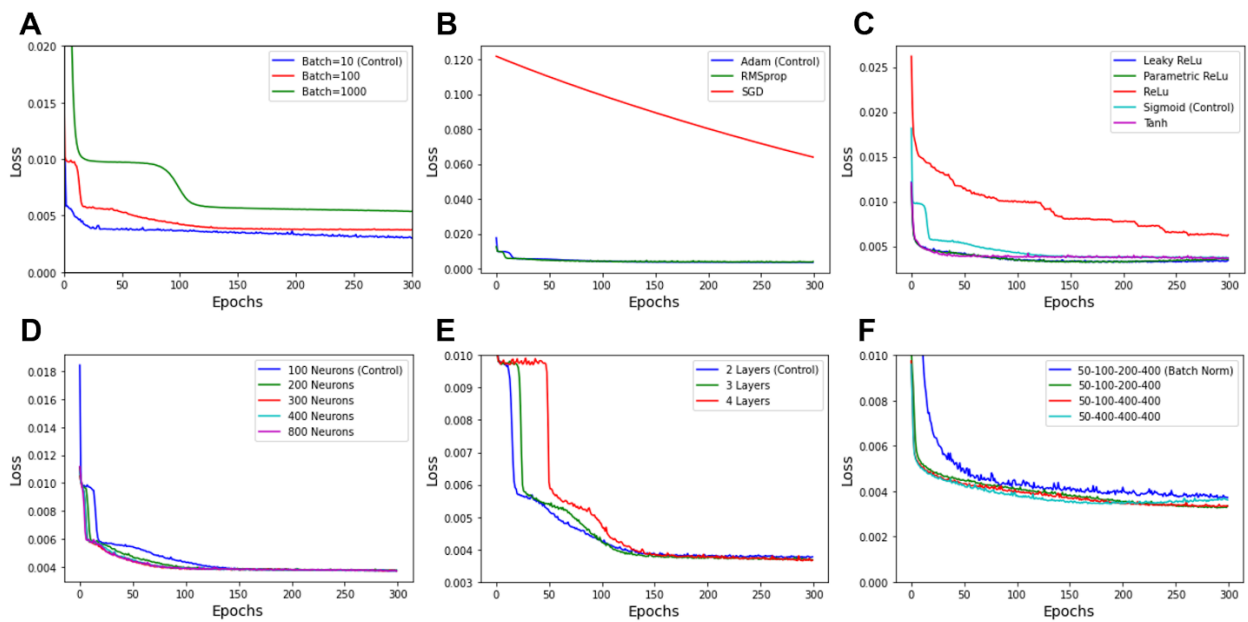

**Figure S2**. Forward dense network hyperparameter tuning. Validation loss of a controlled architecture in comparison to various dependent variables, including: (A) batch size, (B) optimizer, (C) activation function, (D) number of neurons per layer, and (E) number of hidden layers. (F) Final optimized architecture comparison.

Next, we evaluate the performance of the plain 1-D CNN for supercell design. Similar to the dense networks tested in the previous section, we trained multiple CNNs with architecture depths ranging from two to four 1-D convolutional layers (Figure S3A). Each convolutional layer is composed of 64 filters with a kernel size of 3, followed by Leaky ReLU and batch normalization layers and a dense output layer. From these tests, we observe that the plain CNNs yield lower losses than the dense networks. We then converted the plain CNN into the ResNet architecture, which was previously reported to outperform plain CNNs of the same depth. In particular, ResNets address the vanishing-gradient problem using identity shortcuts or skipped connections to form "residual blocks". The residual blocks take advantage of additional identity functions to allow the smooth forward and backward propagation of gradients. As seen in Figure S3B, we trained multiple ResNets with various numbers of residual blocks. Each residual block within the ResNet contains two convolutional layers with other hyperparameters identical to those of the plain CNN. We observe that a two-block model (with four total layers) is capable of outperforming the plain CNN with the same amount of layers. When tuning the other hyperparameters, we found that the ReLU activation (Figure S3C) with a batch size of 10 (Figure S3D) and 32 filters (Figure 3E) yielded the best performance. Figure S3F presents a final comparison between the optimized dense, plain CNN, and ResNet architectures, where it can be observed that the ResNet possesses the lowest validation loss of approximately $8.0 \times 10^{-4}$.

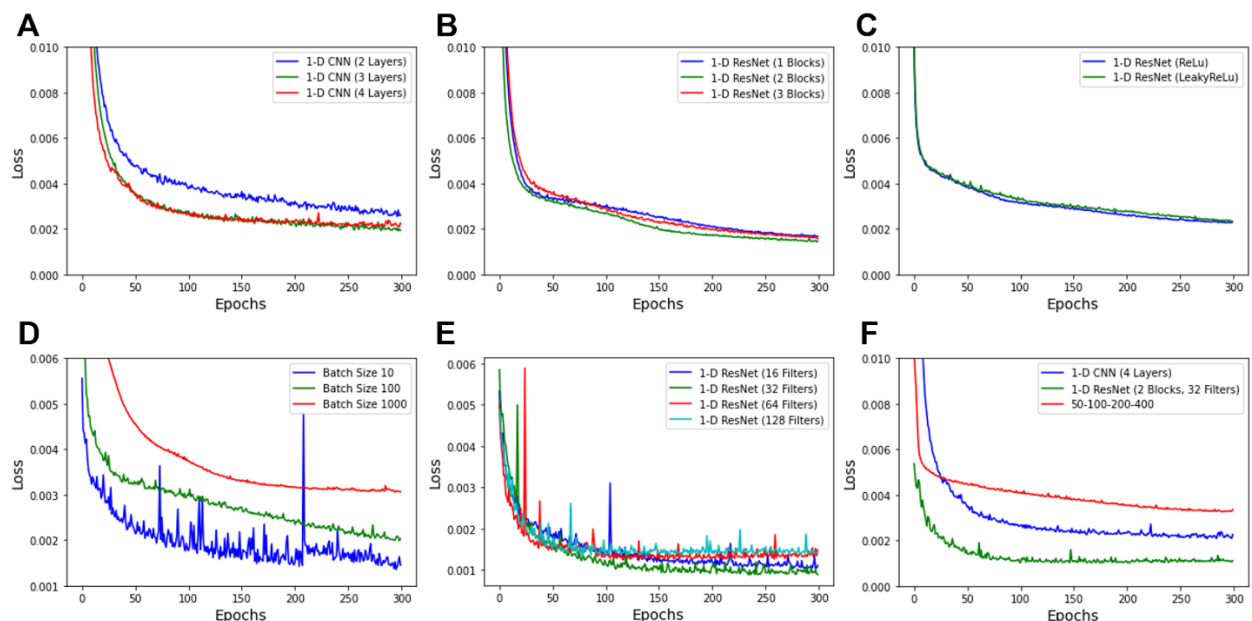

**Figure S3**. Forward CNN and ResNet hyperparameter tuning. Optimization of the number of (A) convolutional layers in a plain CNN and (B) residual blocks in a ResNet. Further tuning results of the (C) ResNet activation function, (D) batch size, (E) and number of filters per layer. Final comparison between the optimized dense, plain CNN, and ResNet architectures.

**Nonuniqueness Analysis**

The primary advantage of the tandem network architecture is its ability to address the nonuniqueness scattering problem (as described in the main text), where drastically different design parameters can meet a similar target response (*i.e.*, a one-to-many problem). To illustrate this problem in the context of our training dataset, Figure S4 shows examples where two substantially different supercell design configurations map to nearly-identical dual-band (Figure S4A) and triple-band (Figure S4B) responses. In Figure S5A, the nonunique dual-band and triple-band spectra were passed into an inverse modeling network, and a poor match between the input spectra and the simulated spectra of the optimized design is observed. However, when the same spectra were passed into the tandem network (Figure S5B), the match between the input spectra and the simulated parameters is substantially improved. Thus, we validate that the tandem network is better than the inverse modeling network at resolving the nonuniqueness issue for the explored class of metasurfaces.

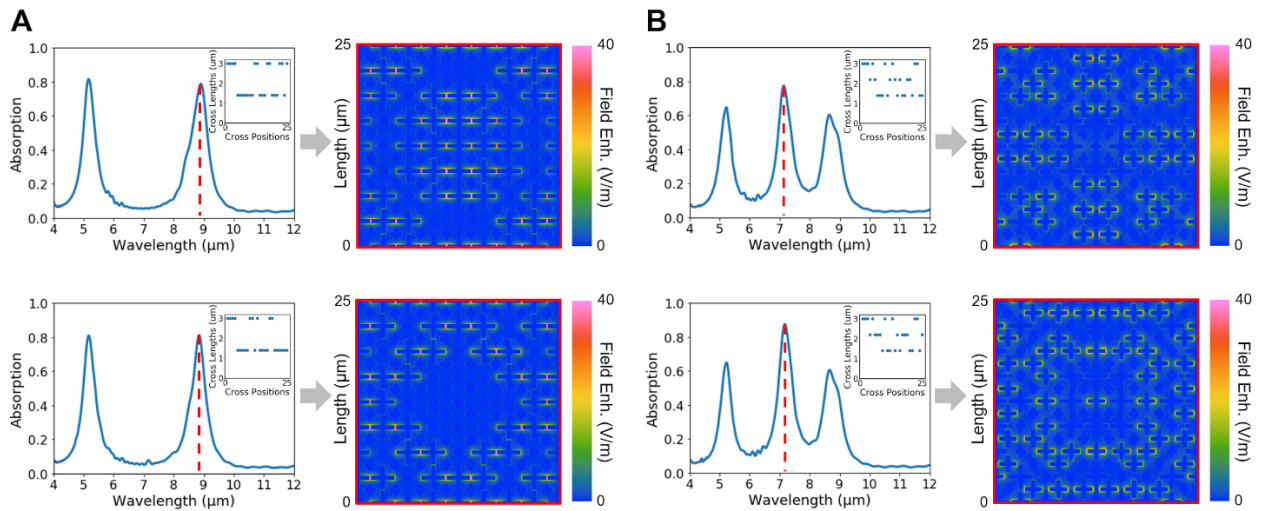

**Figure S4.** Examples of nonuniqueness in the training dataset. Two very different supercell design layouts have nearly-identical (A) dual-band and (B) triple-band absorption responses. Inset images show the design parameters corresponding to the shown absorption spectra.

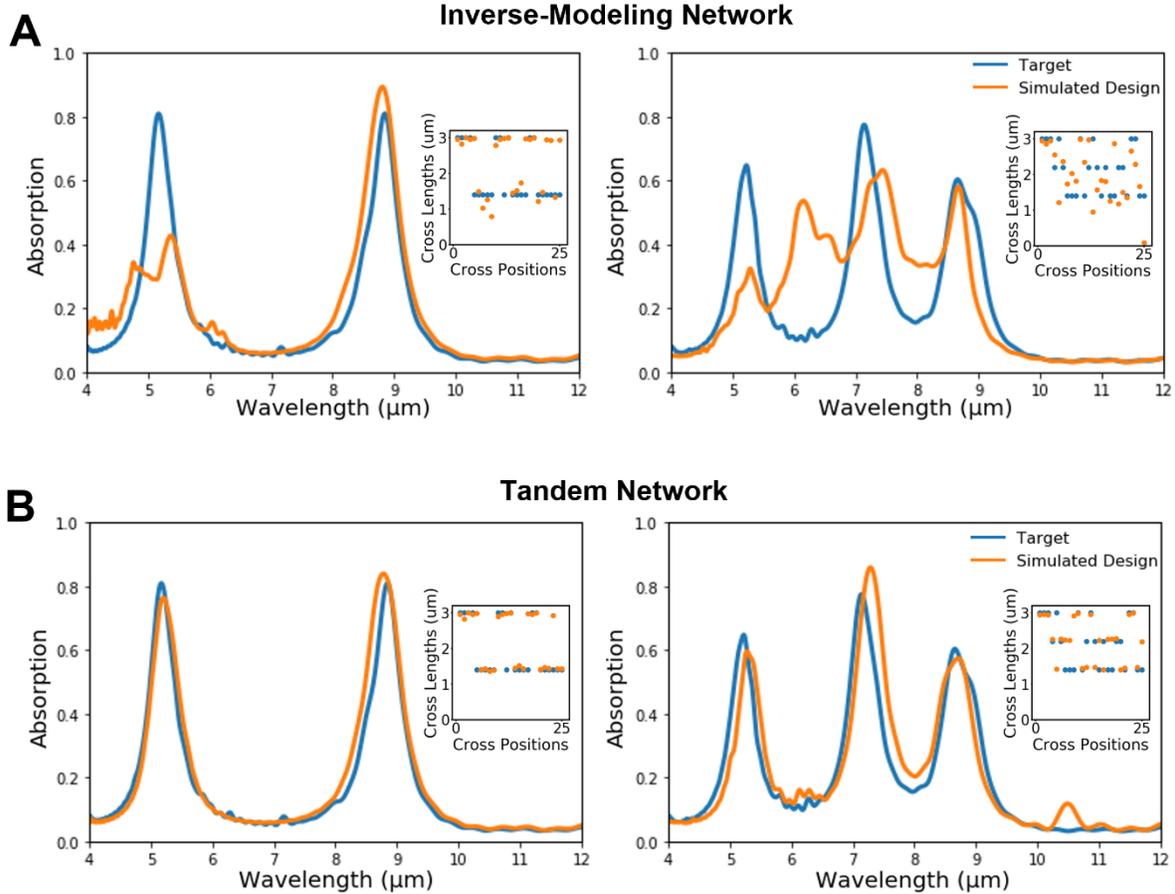

**Figure S5.** Comparisons between target absorption spectra (from the previous Figure) and the simulated spectra for the optimized designs from the (A) inverse-modeling network and the (B) tandem network. The tandem network has greater accuracy in predicting design parameters than the inverse network due to stronger convergence. Inset images show design parameters for the corresponding spectra.

**Training Dataset Size Analysis**

The number of supercell design parameters in *D* (25 total resonators) and the range of values therein (9 different cross lengths) mean there are a total of $3.81 \times 10^{12}$ possible supercell designs in the investigated design space. To expedite our deep learning efforts, we sought to minimize the total simulation time and therefore the amount of training data. However, we also had to ensure that the dataset size was large enough to maximize the network's ability to learn supercell designs. In that regard, as shown in Figure S6A, we trained the optimized tandem network architecture with various quantities of training data and evaluated the corresponding validation losses. We recorded the network's final validation loss using two different validation datasets. The first validation dataset was a fixed group of 300 data instances. This dataset was intended to monitor the network's growth as it trained toward a predetermined set of goals. The second validation dataset came from randomly splitting 10% of the data instances within the available training set. The validation loss derived from the second dataset informs how well the network is able to generalize from the amount of data it learned from.

Increasing increments of randomly-sampled supercell cross lengths, where $D$ contained values ranging from 1.4-3 µm, were initially generated and used for training. However, after training over 1,500 designs, intermediate tests revealed that the network struggled to predict spectrally selective designs with high absorption (as seen in Figure S6B). Instead, the network was biased towards producing quasi-broadband responses as a result of the dispersed distribution of cross lengths in both the training dataset and the network's predictions. To address this shortcoming, we tailored the distribution of cross length values to enforce spectral selectivity. Additional training data were generated by restricting the cross length ranges through a pseudo-random sampling technique, in which $D$ only contains values from discrete ranges (1.4-1.8 µm, 2.0-2.4 µm, etc.) rather than the entire permissible range. Nonsequential values and ranges were also included in this pseudo-random sampling process (where $D$ only contains cross lengths of 1.4-1.6 µm and 2.8-3.0 µm, or 1.8-2.0 µm and 2.6-2.8 µm, and so forth). After including the pseudo-randomly generated designs, it can be observed (in Figure S6C) that the predictions of spectrally selective designs improved considerably.

We observe that both validation losses converge to $2.5 \times 10^{-3}$ at 3,600 data instances, demonstrating that the final model resulted in the optimal performance while minimizing the amount of data required for deep learning by sampling less than $1 \times 10^{-7}$ % of the total design space. On a distributed high-performance computing cluster with four dedicated compute nodes per simulation, where a node has a minimum of four 64-bit Intel Xeon or AMD Opteron CPU cores and 8 GB memory, each FDTD simulation took approximately 30 minutes to complete. Therefore, our training dataset equates to approximately 75 days of simulation time.

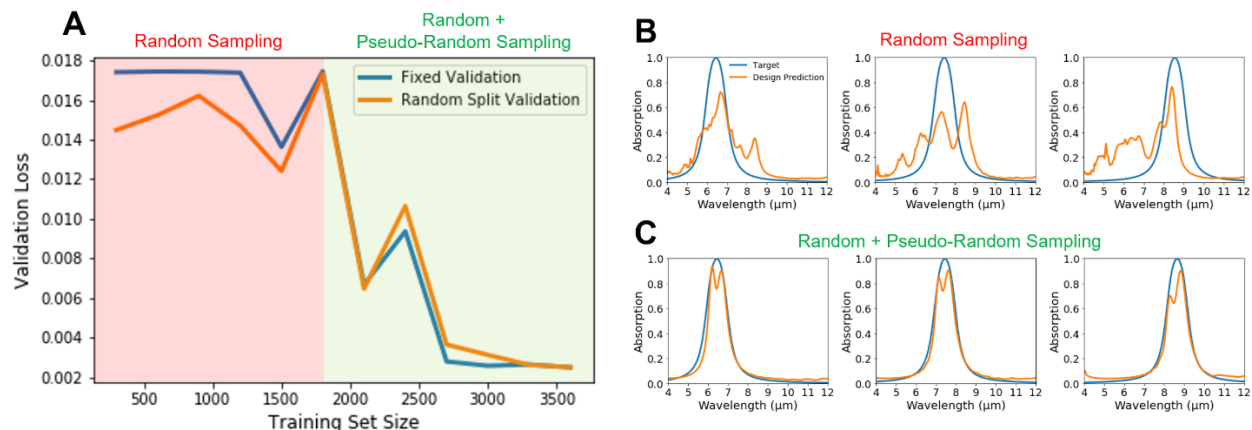

**Figure S6.** Training set size analysis. (A) Validation loss vs. training set size with fixed and randomly split validation datasets. The training set consists of both randomly sampled and pseudo-randomly sampled designs. Test results of a network trained on (B) only randomly sampled designs and (C) both randomly sampled and pseudo-randomly sampled designs.